\documentclass[sigconf]{acmart}
\AtBeginDocument{%
  }

\copyrightyear{2026}
\acmYear{2026}
\setcopyright{cc}
\setcctype{by}

\usepackage{subfig}
\usepackage{graphicx}
\usepackage{multirow}
\usepackage{array}
\usepackage{bbm}

\begin{document}

\title[Meta-Learning and Targeted Differential Privacy to Improve the Accuracy--Privacy Trade--Off in Recommendations]{Meta-Learning and Targeted Differential Privacy to Improve the Accuracy--Privacy Trade--off in Recommendations}

\author{Peter Müllner}
\orcid{0000-0001-6581-1945}
\affiliation{%
  \institution{Know Center Research GmbH}
  \city{Graz}
  \country{Austria}}
\email{pmuellner@know-center.at}

\author{Dominik Kowald}
\orcid{0000-0003-3230-6234}
\affiliation{%
  \institution{Know Center Research GmbH \& University of Graz}
  \city{Graz}
  \country{Austria}
}
\email{dkowald@know-center.at}

\author{Markus Schedl}
\orcid{0000-0003-1706-3406}
\affiliation{%
  \institution{Johannes Kepler University \& LIT Linz}
  \city{Linz}
  \country{Austria}}
\email{markus.schedl@jku.at}

\author{Elisabeth Lex}
\orcid{0000-0001-5293-2967}
\affiliation{%
  \institution{Graz University of Technology}
  \city{Graz}
  \country{Austria}}
\email{elisabeth.lex@tugraz.at}


\begin{abstract}
  Balancing differential privacy (DP) with recommendation accuracy is a key challenge in privacy-preserving recommender systems, since DP-noise degrades accuracy.
  We address this trade-off at both the data and model levels. 
  At the data level, we apply DP only to the most stereotypical user data likely to reveal sensitive attributes, such as gender or age, to reduce unnecessary perturbation; we refer to this as \emph{targeted DP}.
  At the model level, we use meta-learning to improve robustness to remaining DP-noise.
  This achieves a better trade-off between accuracy and privacy than standard approaches: 
  Meta-learning improves accuracy and targeted DP leads to lower empirical privacy risk compared to uniformly applied DP and full DP baselines.
  Overall, our findings show that selectively applying DP at the data level together with meta-learning at the model level can effectively balance recommendation accuracy and user privacy.
\end{abstract}




\maketitle

\begin{figure}[t]
    \centering
    \includegraphics[width=0.9\linewidth]{approach.pdf}
    \caption{Unlike methods that apply DP uniformly, we propose a two-stage approach: at the data level, DP is applied only to strongly stereotypical data (targeted DP) to reduce unnecessary noise, and at the model level, meta-learning improves robustness to residual DP noise. This improves the accuracy–privacy trade-off over standard methods.}
    \label{fig:intro}
\end{figure}

\section{Introduction and Background}
Recommender systems are central to many online platforms, filtering content in domains such as e-commerce, media streaming, and social networks. 
They rely heavily on user interaction data, which often exhibits strong stereotypical patterns, i.e., users with similar, potentially sensitive attributes tend to interact with similar types of items. 
While these patterns enable accurate recommendations, they can also expose sensitive user information~\cite{escobedo2024making}.
To prohibit this, differential privacy (DP) has emerged as standard for protecting user data. 
By injecting random noise into data or model updates, DP limits the ability to infer information about individual users, but at the cost of degrading recommendation accuracy. 
Balancing this trade-off between privacy and accuracy remains a key challenge.

In this work, we address this by separating the source of DP-noise at the data level from its impact at the model level, designing complementary solutions for both.
Traditional approaches apply DP uniformly across all user data~\cite{mcsherry2009differentially}, introducing unnecessary perturbation and accuracy loss, even though only stereotypical data is likely to reveal sensitive attributes. 
Prior work has explored applying DP only to subsets of users while leaving others unprotected~\cite{xin2014controlling}. 
In contrast, targeted DP applies DP only to the most stereotypical parts of user data, limiting the injected DP-noise while maintaining DP for the most sensitive data.
Targeted DP uses a data budget $\beta$, leaving a fraction $\beta$ of each user's data unprotected while applying DP to the most stereotypical $1-\beta$, enabling control over accuracy, formal DP guarantees, and empirical privacy risk.
At the model level, residual DP-noise can still affect learning even with targeted DP, therefore, we leverage meta-learning, which enhances adaptation under noisy conditions~\cite{wang2020training} and has proven effective for data scarcity in recommender systems~\cite{lee2019melu,muellner2021robustness}.

Our results show that this combination leads to an improved empirical accuracy–privacy trade-off, maintaining recommendation accuracy while reducing empirical privacy risk compared to standard DP baselines, all while maintaining formal DP guarantees for the most sensitive data.
Overall, our work suggests the importance of addressing the accuracy–privacy trade-off at two levels: controlling the source of DP-noise at the data level with targeted DP, and mitigating its impact at the model level with meta-learning.



\section{Approach}
\subsection{Targeted Application of DP}
We introduce a data budget $\beta \in [0, 1]$, representing the fraction of a user's data $D_u$ that can be used directly without DP. 
The remaining fraction $1-\beta$ is protected with DP before training.

For each user $u$, we identify the fraction of $1-\beta$ most stereotypical data $D^s_u \subseteq D_u$, i.e., item ratings in $u$'s profile.
We follow Escobedo et al.~\cite{escobedo2024making} and measure an item's stereotypicality based on how strongly it indicates a sensitive attribute $a$, i.e., the item-group-inclination $IGI(i, a) = |\{ u \in U_a: r_{u, i} \in D \}| / |U_a|$, where $U_a$ is the set of users with attribute $a$ and $r_{u, i}$ is the rating of user $u$ and item $i$. Formally:
\begin{equation}
    D^{s}_u = \arg\max^k_{r_{u, i} \in D_u} \mathbbm{1}_a(u) \frac{IGI(i, a) - IGI(i, \overline{a})}{\max \{ IGI(i, a), IGI(i, \overline{a}) \}}
    \label{eq:ister}
\end{equation}
where $k = \lceil (1-\beta) \cdot |D_u| \rceil$ and $\overline{a}$ is the complementary attribute to $a$.
This produces values between -1 and 1 depending on whether $IGI(i, a) > IGI(i, \overline{a})$.
Thus, we define $\mathbbm{1}_a(u) = 1$ if $u \in U_a$, and $\mathbbm{1}_a(u) =-1$ if $u \in U_{\overline{a}}$.
For example, if the user has sensitive attribute ``blue'', items in the user's profile that are indicative of the ``blue'' stereotype could reveal the sensitive attribute and thus, we apply DP to these items (cf. Figure~\ref{fig:intro}).
In the remainder, we refer to this strategy as $Targeted_{DP}$, whereas the $Random_{DP}$ baseline samples data points uniformly at random, i.e., $D^s_u \sim \mathrm{Uniform}_k(D_u)$.

To ensure DP, we apply a coin-flip mechanism $m_{DP}$~\cite{chen2025data} to each rating $r_{u, i} \in D^{s}_u$:
With some probability determined by the privacy budget $\epsilon$, $r_{u, i}$ is replaced with a random rating $\tilde{r}_{u, j}$, where the item $j$ is sampled uniformly from items not interacted with by user $u$ and the rating value $\tilde{r}_{u, j}$ is sampled from the set of possible rating values $\mathcal{R}$. 
$\epsilon$ controls how many random items are added:
\begin{equation}
    m_{DP}(r_{u, i}) = \left\{
    \begin{array}{l >{\hspace{1cm}\displaystyle}l}
        r_{u, i} & \textrm{ with probability } \frac{e^\epsilon}{e^\epsilon + 1}, \\
        \tilde{r}_{u, j} & \textrm{ with probability } \frac{1}{e^\epsilon + 1}.
    \end{array}
    \right.
    \label{eq:dp}
\end{equation}
where $j \sim \mathrm{Uniform}(I\setminus I_u)$ and $\tilde{r}_{u, j} \sim \textrm{Uniform}(\mathcal{R})$. 
The final training dataset $\tilde{D}_u$ combines unprotected with DP-protected data:
\begin{equation}
    \tilde{D}_u = (D_u \setminus D^s_u) \cup \{m_{DP}(r_{u, i}): r_{u, i} \in D^s_u\}
\end{equation}
This procedure allows a targeted application of DP to the fraction $1-\beta = |D^s_u|$ of each user's training data while leaving the remaining fraction $\beta = |D_u \setminus D^s_u|$ fully available for the recommender system.

\subsection{Meta-Learning Recommender System}
To generate recommendations, we use MetaMF~\cite{lin2020meta}, i.e., a meta-learning matrix factorization model that incorporates a meta-network to generate user-specific parameters. 
The meta-network enables the model to quickly adapt to individual users' preferences by leveraging shared patterns across all users. 
In contrast, NoMetaMF~\cite{muellner2021robustness} is a variant where meta-learning is disabled by prohibiting backpropagation through the meta-network.
Overall, these models allow us to study the impact of meta-learning on the model's ability to preserve recommendation accuracy under DP.
 
In the remainder, we use $MetaMF$ to denote the baseline without DP and $MetaMF_{DP}$ is the full DP baseline in which all training data is protected with DP. 
Furthermore, $MetaMF_{DP}^{Random}$ and $MetaMF_{DP}^{Targeted}$ refer to models trained on data where only randomly selected or the most stereotypical parts are protected with DP.
The same notation applies to $NoMetaMF$.

\section{Experimental Setup}
\begin{table}[t]
    \resizebox{\linewidth}{!}{
    \centering
    \begin{tabular}{l r r r r r r}
    \toprule
    Dataset & Users & Items & Ratings & R/U & R/I & Density \\ \midrule
    MovieLens 1M & 6,040 & 3,706 & 1,000,209 & 165.50 & 269.89 & 4.47\% \\ 
    Bookcrossing & 3,971 & 5,662 & 65,111 & 16.40 & 11.50 & 0.29\% \\ \bottomrule
    \end{tabular}}
    \caption{Descriptive statistics of the two datasets.}
    \label{tab:datasets}
\end{table}

\subsection{Datasets}
For our experiments, we use the \emph{MovieLens 1M}~\cite{harper2015movielens} and \emph{Bookcrossing}~\cite{ziegler2005improving} datasets. 
Due to the high sparsity of Bookcrossing, we apply 5-core user- and item pruning and remove implicit ratings~\cite{naghiaei2022unfairness}.
Descriptive statistics for both datasets are summarized in Table~\ref{tab:datasets}.
Furthermore, for MovieLens 1M, we use gender (28\% female, 72\% male) as sensitive attribute, while for Bookcrossing, we bin users into two age groups using the median age (32) as the threshold (after pruning: 41\% are younger and 59\% are 32 or older).

\subsection{Attack Model}

For the attacker, we employ a 2-layer feedforward neural network with ReLU activations and two linear output neurons~\cite{escobedo2024making}.  
The input consists of users' rating vectors, where unrated items are set to 0 and rated items are represented by their corresponding rating values.  
The attacker's objective is to infer users' sensitive attributes, i.e., gender for {MovieLens 1M} and age group for {Bookcrossing}.


\subsection{Evaluation Protocol}
For each $(\beta, \epsilon)$-pair, we generate a 60/20/20 train/validation/test split and train and evaluate recommendation models across five folds. 
We use $\beta \in \{1, 0.9, 0.8, \dots, 0.1, 0\}$ and $\epsilon \in \{3, 2, 1, 0.1\}$.
Hyperparameters are tuned for each $\epsilon$-value, $\beta=0$ (all data protected with DP), and $\beta=1$ (no DP applied).
The hyperparameters with the lowest mean absolute error on the validation set are used for evaluation\footnote{The exact parameter values as well as the source code can be found in our publicly available repository~\url{https://github.com/pmuellner/MetaTargetedDP}.}.
For the attacker, we perform 10 runs, each with a random 60/20/20 split.
Hyperparameters are tuned for all $(\epsilon, \beta)$-pairs in the first run and then kept the same for all runs.

\paragraph{Metrics}
We measure recommendation accuracy using the mean absolute error ($MAE$), and quantify the impact of DP via $\Delta MAE$, i.e., the change relative to the corresponding non-DP baseline. 
To assess privacy, we use balanced accuracy ($BAcc$) to measure an attacker’s ability to infer sensitive attributes and report $\Delta BAcc$ as the corresponding change. 
While $\Delta BAcc$ captures empirical privacy risk, formal guarantees are provided by the privacy budget $\epsilon$.

\section{Results and Discussion}
\begin{table}[t]
    \centering
    \resizebox{\linewidth}{!}{
    \begin{tabular}{l r r r r r r r r r r}
    \toprule
    & & &\multicolumn{4}{c}{MovieLens 1M} & \multicolumn{4}{c}{Bookcrossing} \\ \cmidrule(lr){4-7} \cmidrule(lr){8-11}
    & $\epsilon$ & $\beta$ & $MAE$ & $\Delta MAE$ & $BAcc$ & $\Delta BAcc$ & $MAE$ & $\Delta MAE$ & $BAcc$ & $\Delta BAcc$ \\ \midrule
    {$MetaMF$} & $\infty$ & 1.0 & 0.6857 & 0.00\% & 0.7429 & 0.00\% & 1.1789 & 0.00\% & 0.6582 & 0.00\% \\ \midrule
    \multirow{4}{*}{$MetaMF_{DP}$} & 3.0 & 0.0 & 0.6985 & 1.87\% & 0.7438 & 0.12\% & 1.1821 & 0.27\% & 0.6498 & -1.27\%\\
    & 2.0 & 0.0& 0.7027 & 2.49\% & 0.7390 & -0.53\% & 0.7387 & -37.34\% & 0.6630 & 0.74\%\\
    & 1.0 & 0.0& 0.7216 & 5.25\% & 0.7271 & -2.12\% & 0.9050 & -23.23\% & 0.6397 & -2.81\%\\
    & 0.1 & 0.0& 0.7568 & 10.37\% & 0.7176 & -3.40\% & 1.5039 & 27.57\% & 0.6419 & -2.48\%\\ \midrule \midrule
    {$NoMetaMF$} & $\infty$ & 1.0 & 0.7009 & 0.00\% & 0.7429 & 0.00\% & 1.3484 & 0.00\% & 0.6582 & 0.00\%\\ \midrule
    \multirow{4}{*}{$NoMetaMF_{DP}$} & 3.0 & 0.0 & 0.7072 & 0.90\% & 0.7438 & 0.12\% & 1.3781 & 2.20\% & 0.6498 & -1.27\%\\
    & 2.0 & 0.0 & 0.7227 & 3.10\% &  0.7390 & -0.53\% & 0.8270 & -38.67\% & 0.6630 & 0.74\%\\
    & 1.0 & 0.0 & 0.7450 & 6.29\% & 0.7271 & -2.12\% & 1.0833 & -19.66\% & 0.6397 & -2.81\%\\
    & 0.1 & 0.0 & 0.7767 & 10.80\% & 0.7176 & -3.40\% & 2.0971 & 55.52\% & 0.6419 & -2.48\% \\ \bottomrule
    \end{tabular}}
    \caption{Evaluation results for $MetaMF$ and $NoMetaMF$, i.e., no DP, and $MetaMF_{DP}$ and $NoMetaMF_{DP}$, i.e., all data protected. Overall, we find that meta-learning provides the lowest $MAE$ and can slightly reduce the loss in recommendation accuracy when DP is applied, especially for small $\epsilon$-values. Plus, the empirical privacy risk decreases when stronger formal DP guarantees are used, i.e., smaller $\epsilon$-values.}\vspace{-7mm}
    \label{tab:baselines}
\end{table}

\subsection{Meta-Learning under DP}
\label{subsec:meta_learning_and_dp}

As the $MAE$-values in Table~\ref{tab:baselines} show, $MetaMF_{DP}$ consistently outperforms $NoMetaMF_{DP}$ across all privacy budgets. 
For high $\epsilon$ (weak DP), both methods show minimal degradation as measured by $\Delta MAE$; however, this degradation is slightly smaller for $MetaMF_{DP}$ than for $NoMetaMF_{DP}$. 
For MovieLens 1M, as $\epsilon$ decreases (stronger DP), the $MAE$ increases for both models. 
On Bookcrossing, DP ($1 \leq \epsilon \leq 2$) acts as a regularizer, improving generalization. 
As $\epsilon$ decreases further, the $MAE$ increases again, with $NoMetaMF_{DP}$ experiencing a larger increase compared to $MetaMF_{DP}$. 

Furthermore, the $\Delta BAcc$-values for both datasets indicate that lower $\epsilon$-values correspond to lower empirical privacy risk as measured by $\Delta BAcc$. 
Here, there are no differences between $MetaMF_{DP}$ and $NoMetaMF_{DP}$, since $\Delta BAcc$ reflects the attacker and is not a property of the recommender system.

Overall, our experiments show that stronger DP leads to lower recommendation accuracy due to the noise introduced by DP, and that meta-learning systematically reduces this degradation, enhancing robustness against DP-noise.


\subsection{Targeted Application of DP}
\label{subsec:accuracy}

\begin{figure}
    \centering
    \includegraphics[width=\linewidth]{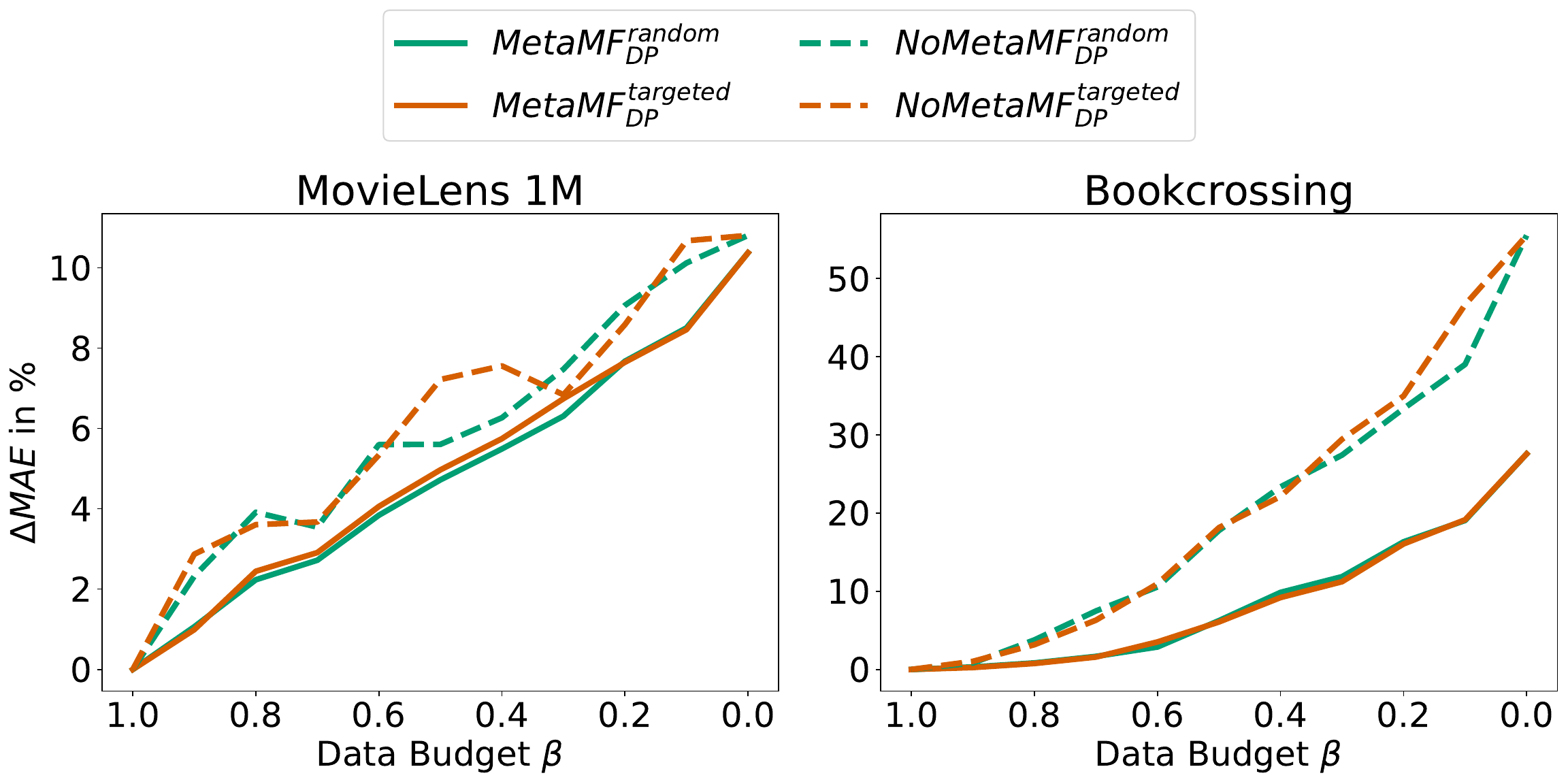}
    \caption{$\Delta MAE$ for random and targeted application of DP ($\epsilon=0.1$). As $\beta$ decreases, more data is protected, and $\Delta MAE$ increases. $MetaMF$ outperforms $NoMetaMF$, highlighting the advantage of meta-learning. Targeted DP achieves similar $\Delta MAE$ as random application, showing that stereotypical data can be protected without harming recommendation accuracy.}
    \label{fig:targeted_dp_mae}
\end{figure}

The results in Figure~\ref{fig:targeted_dp_mae} show an increase in $\Delta MAE$ as $\beta$ decreases, as larger parts of each user's data are protected with DP (we report results for $\epsilon = 0.1$ and $\epsilon=3$).
Furthermore, no systematic differences in the loss of recommendation accuracy between random selection and selecting the most stereotypical items can be observed.
Thus, stereotypical data can be protected without additional accuracy loss. 
However, $MetaMF$ consistently exhibits lower accuracy degradation than $NoMetaMF$, supporting our findings from Section~\ref{subsec:meta_learning_and_dp}. 
This is particularly pronounced for Bookcrossing as $\beta$ decreases.

\begin{figure}[!t]
    \centering
    \includegraphics[width=\linewidth]{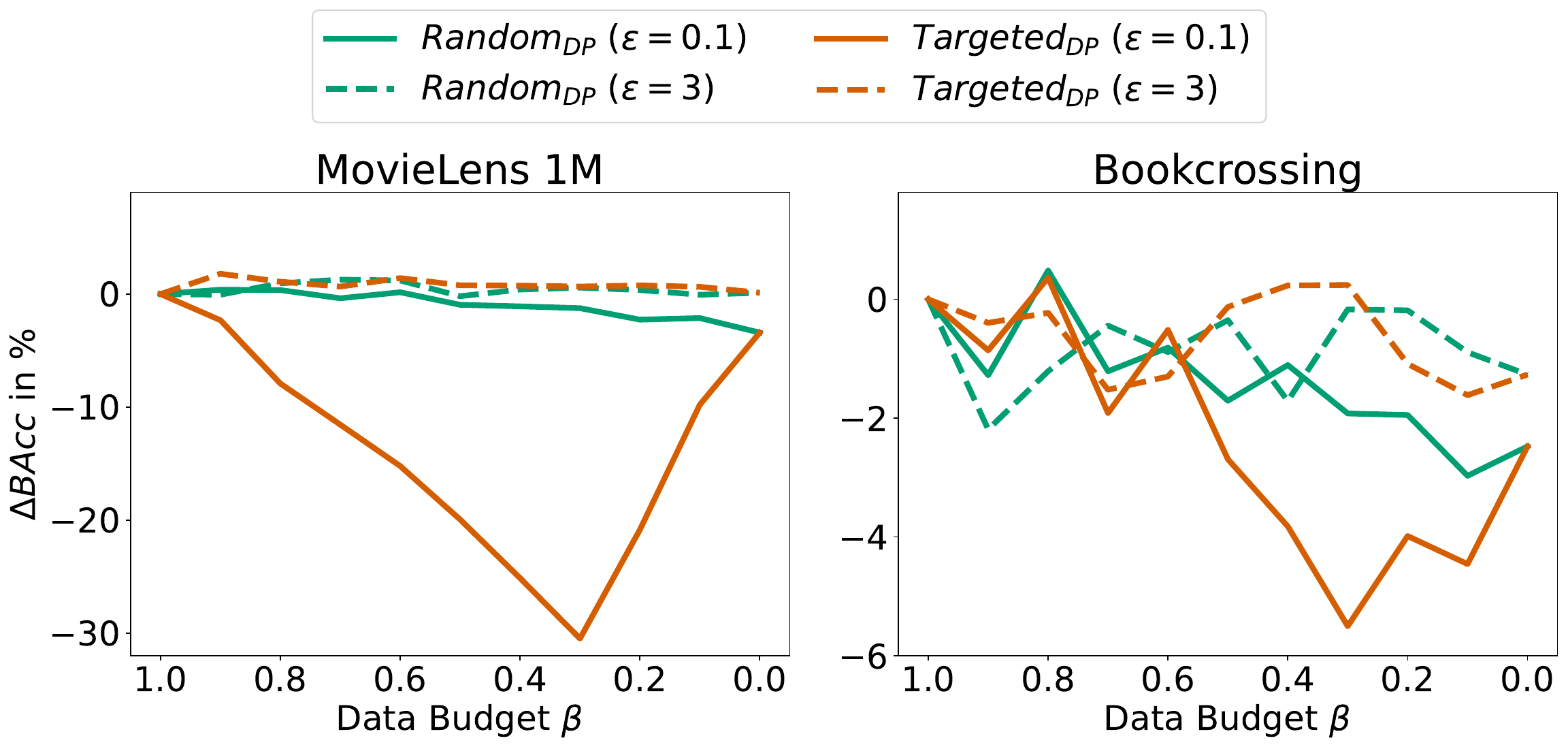}
    \caption{$\Delta BAcc$ for random and targeted application of DP. Weak DP ($\epsilon = 3$) has little effect, while strong DP guarantees ($\epsilon = 0.1$) with targeted DP effectively reduces $\Delta BAcc$. 
    At $\beta=0.3$, users reach ``neutral'' stereotypicality and the lowest $\Delta BAcc$, even below the full DP baseline ($\beta = 0$), highlighting the impact of $\beta$ on empirical privacy risk.}
    \label{fig:targeted_dp_bacc}
\end{figure}

Moreover, as Figure~\ref{fig:targeted_dp_bacc} shows, $Targeted_{DP}$ effectively reduces empirical privacy risk measured by $\Delta BAcc$, especially under stricter privacy regimes ($\epsilon = 0.1$). 
It also yields much lower $\Delta BAcc$ compared to $Random_{DP}$ and even the full DP baseline ($\beta = 0$). 
For both datasets, $\Delta BAcc$ reaches a minimum at $\beta = 0.3$ and then increases for lower values:
As $\beta$ decreases, more stereotypical items are replaced by random items due to the DP-mechanism.
These replacements could be stereotypical or antistereotypical depending on the dataset's stereotypicality distribution\footnote{For MovieLens 1M, 46.7\% of items are indicative of the female gender, whereas 46.2\% are indicative of the male gender. 
For Bookcrossing, 46.2\% of items indicate a user age under 32, while 44.8\% indicate an age of 32 or older. Measured via Equation~\ref{eq:ister}.}.
Therefore, stereotypical items are replaced by antistereotypical items, leading to a more ``neutral'' user stereotypicality, making the user data less likely to reveal the sensitive attribute, which results in lower $\Delta BAcc$.
For $\beta=0.3$, the average user stereotypicality is at a minimum of 0.04 (0.13 without DP) for MovieLens 1M and 0.12 for Bookcrossing (0.25 without DP), showing that sensitive attributes are weakly reflected in user representations, which makes attribute inference particularly difficult.
Even though both datasets exhibit very different descriptive statistics (see Table~\ref{tab:datasets}), the average fraction of stereotypical items per user profile is similar (i.e., 60\% for MovieLens 1M and 61\% for Bookcrossing).
At $\beta = 0.3$, all stereotypical items are protected with DP, creating a similar ``sweet spot'' for both datasets.
For $\beta < 0.3$, DP is also applied to many antistereotypical items and their replacements increase user stereotypicality.

Overall, these results indicate that selectively applying DP to the most stereotypical data protects sensitive information without additional accuracy loss, particularly when combined with meta-learning at the model level. 
Furthermore, there exists a sweet spot (in our case $\beta = 0.3$), where users achieve ``neutral'' stereotypicality, minimizing empirical privacy risk, while keeping the most sensitive and stereotypical user data under formal DP guarantees.

\subsection{Discussion}
\begin{figure}[!t]
    \centering
    \includegraphics[width=\linewidth]{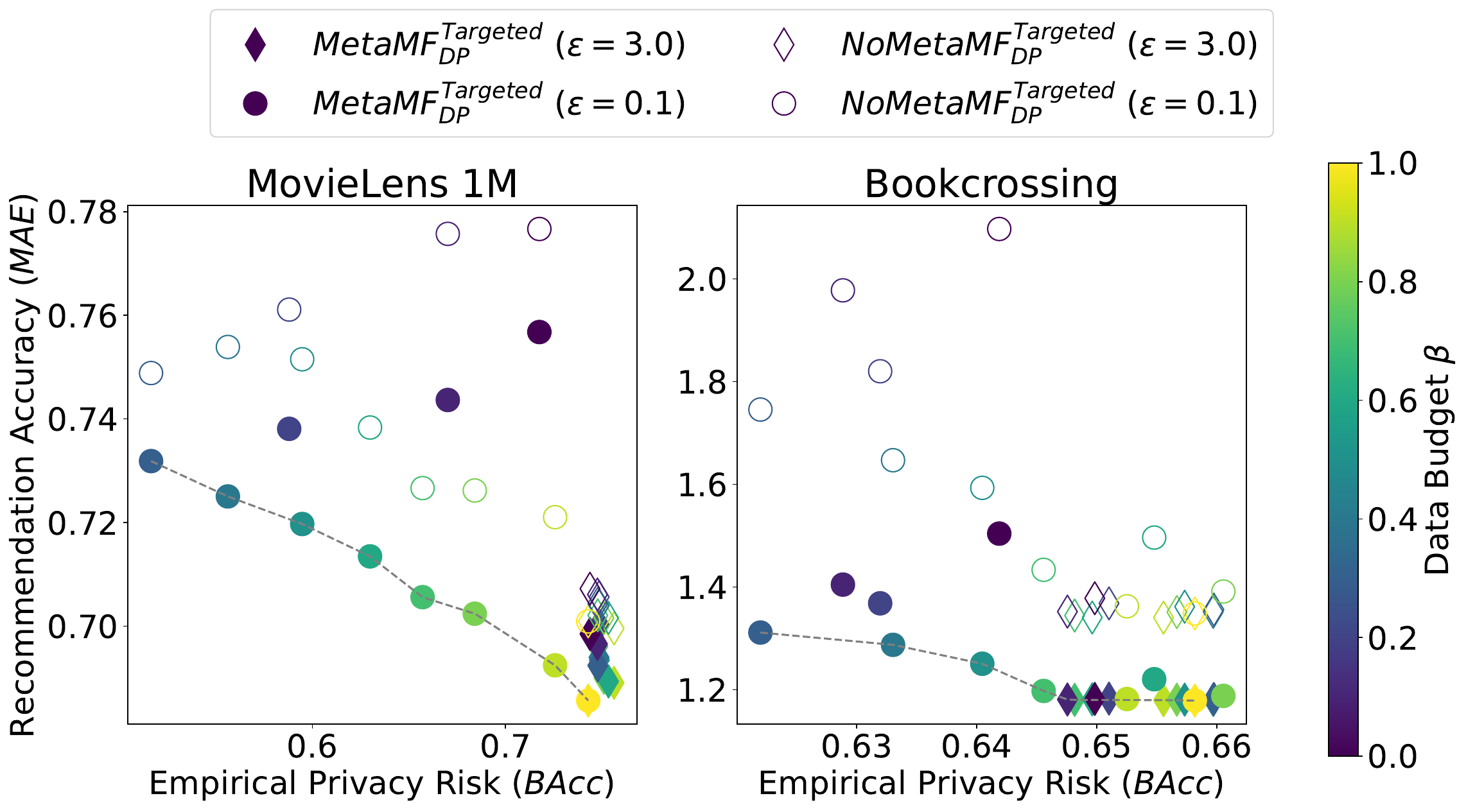}
    \caption{Comparison of $MetaMF$ and $NoMetaMF$ under targeted DP with varying $\epsilon$-values. The dashed line shows the Pareto frontier over $BAcc$ and $MAE$. Only $MetaMF$ lies on the frontier, demonstrating the robustness of meta-learning to DP-noise. Fully protecting all data ($\beta=0$) is suboptimal; adjusting the data budget $\beta$ provides a lever to balance accuracy and empirical privacy risk in practice.}
    \label{fig:trade-off}
\end{figure}
As illustrated by the Pareto frontier in Figure~\ref{fig:trade-off}, $MetaMF$ consistently achieves a better trade-off between recommendation accuracy and empirical privacy risk than $NoMetaMF$ in our scenario, particularly in strict privacy regimes ($\epsilon=0.1$). 
While $NoMetaMF$'s accuracy degrades fast as DP-noise increases, $MetaMF$ maintains accuracy, showing that meta-learning helps mitigate the impact of DP-noise on the model level. 

On the data level, Figure~\ref{fig:targeted_dp_mae} and Figure~\ref{fig:targeted_dp_bacc} already revealed that targeted DP leads to similar recommendation accuracy, but lower empirical privacy risk than a random application of DP.
Therefore, we now focus on the trade-off of targeted DP and the influence of $\beta$: 
The data budget $\beta$ controls the trade-off between recommendation accuracy, empirical privacy risk, and formal DP guarantees. 
While $\beta = 0$ provides full DP guarantees, it does not necessarily minimize empirical privacy risk. 
Intermediate values (e.g., $\beta = 0.3$) achieve the best trade-off, protecting $70\%$ of each user’s most stereotypical data while maintaining moderate accuracy loss.

Overall, these results demonstrate that achieving an effective trade-off requires addressing the impact of DP-noise at the model level through meta-learning and mitigating its source at the data level via targeted DP.
For this purpose, the data budget provides a lever to balance between recommendation accuracy, empirical privacy risk, and formal DP guarantees.




\paragraph{Current Limitations.}

Our findings depend on the attacker used to estimate empirical privacy risk, and results may vary under alternative attacks.
Also, we focus solely on matrix factorization, so the generalizability of our results to other recommendation models remains uncertain.
Computing item stereotypicality further requires access to sensitive attributes during preprocessing, which may introduce additional privacy risks.

\section{Conclusion and Outlook}


This work combines meta-learning at the model level with targeted DP at the data level to improve the accuracy–privacy trade-off. 
Targeted DP protects the most stereotypical user data that could reveal sensitive attributes, while meta-learning reduces accuracy loss caused by DP-noise. 
Our results demonstrate that this combination enhances recommendation accuracy, lowers empirical privacy risk relative to full DP, and preserves formal DP guarantees for the most sensitive data. 
Herein, the data budget $\beta$ serves as a practical lever to balance this trade-off. 
Overall, these findings underscore that effectively managing both the source of DP-noise at the data level and its impact at the model level is essential for accurate and privacy-preserving recommender systems.

Future work will investigate methods to identify high-risk data and design resource-efficient alternatives to meta-learning to further improve robustness under DP-noise.



\begin{acks}
This work was funded in part by the FFG projects ``CareLLM'' (FO999933866) and ``HybridAir'' (FO999902654), and by the FWF (10.55776/COE12).
It was further supported by the project EuroCC Austria which has received funding from the European High Performance Computing Joint Undertaking (JU) and Germany, Bulgaria, Austria, Croatia, Cyprus, Czech Republic, Denmark, Estonia, Finland, Greece, Hungary, Ireland, Italy, Lithuania, Latvia, Poland, Portugal, Romania, Slovenia, Spain, Sweden, France, Netherlands, Belgium, Luxembourg, Slovakia, Norway, Türkiye, Republic of North Macedonia, Iceland, Montenegro, Serbia (101101903). The computational results have been achieved using the Austrian Scientific Computing (ASC) infrastructure.
\end{acks}

\bibliographystyle{ACM-Reference-Format}
\bibliography{bibliography_short}

\end{document}